\documentclass[prd,aps,preprint,tightenlines,superscriptaddress]{revtex4}
\usepackage{epsfig}
\usepackage{axodraw}
\usepackage{amsmath}
\def\euv{ \epsilon_{\text{UV}} }
\def\eir{ \epsilon_{\text{IR}} }

\def\bn{\bar n}

\begin{document}
%\preprint{DOE/ER/40762-274 \ UM-PP\#03-035}

\title{Threshold Resummation for Drell-Yan Process \\ in Soft-Collinear Effective Theory}
\author{Ahmad Idilbi}
\email{idilbi@physics.umd.edu} \affiliation{Department of Physics,
University of Maryland, College Park, Maryland 20742, USA}
\author{Xiangdong Ji}
\email{xji@physics.umd.edu} \affiliation{Department of Physics,
University of Maryland, College Park, Maryland 20742, USA}
\date{\today}
\vspace{0.5in}
\begin{abstract}
We consider Drell-Yan process in the threshold region
$z\rightarrow 1$ where large logarithms appear due to soft-gluon
radiations. We present a soft-collinear effective theory approach
to re-sum these Sudakov-type logarithms following an earlier
treatment of deep inelastic scattering, and the result is
consistent with that obtained through a standard perturbative QCD
factorization.\\

\end{abstract}
\maketitle
\def\euv{ \epsilon_{\text{UV}} }
\def\eir{ \epsilon_{\text{IR}} }

\def\bN{\bar N}

\section{Introduction}

It has been known for many years \cite{alta} that next-to-leading
order (NLO) calculations in perturbative quantum chromodynamics
(pQCD) for cross sections of various hard scattering processes,
lead to functions singular at the edge of the relevant phase
space. For example, for Drell-Yan (DY) process we define $z=Q^2/s$
where $Q^2$ is the invariant mass of the produced lepton pair
squared and $s$ is the total partonic invariant mass squared. At
$n\rm th$ order in pQCD we encounter terms like
$\alpha_s^n[(1-z)^{-1}\ln^{2n-1}(1-z)]$. In the limit
$z\rightarrow1$ these singular functions---when Mellin
transformed---give rise to large double logarithms in moment space
$(\alpha_s \ln^2N)^n$. When $N$ is large, the product of these
large logarithms with a small coupling constant $\alpha_s$ is not
necessarily small and an all-order resummation in perturbation
theory is needed to make predictions reliable. The purpose of a
resummation is to control these large logarithms and to obtain a
closed expression that sums part or all of these logarithms. For
DY and deep-inelastic scattering (DIS) processes, an extensive
theoretical work was performed in
Refs.\cite{sterman8797,catani8991}, according to which, one has to
establish first a factorized form of the cross section into a hard
part (``Wilson coefficient''), jet part and soft part, and then
make a clever use of an evolution in parton rapidity. We will have
more to say about this in section IV.

Recently an effective field theory approach was implemented
\cite{manohar} to re-sum leading logs (LL) and next-to-leading
logs (NLL) for DIS at large momentum transfer and near the
threshold limit $x\rightarrow1$ (where $x$ is the longitudinal
hadron momentum fraction carried by the quark). That analysis was
performed within the soft-collinear effective theory (SCET)
framework \cite{bauer1,bauer2,feldman}. By utilizing this
effective theory for the inclusive DIS (for large momentum
transfer and near threshold) cross sections, one can re-derives
the full QCD results (as a consistency check of the effective
theory!) More importantly, the exponentiation of the LL and NLL in
moment space appears quite naturally in this context following the
solution of a simple RG equation \cite{manohar}.

In this work, we extend the application of SCET to the DY cross
section in the threshold limit $z\rightarrow1$ and show how the
resummation is obtained. We work in the center-of-mass (c.o.m.)
frame of the incoming quark and anti-quark, and identify three
relevant scales: $Q^2 $, $Q^2(1-z)^2$ and $\Lambda_{\rm QCD}^2$.
In the threshold limit we have $1-z\ll1$. Assuming $\Lambda_{\rm
QCD}^2/Q^2\ll(1-z)^2$, we have a clear separation of scales. Given
these three scales one has to perform a two stage matching and
running: First, at $Q^2$ we match the full QCD (annihilation)
current onto the SCET current, giving rise the so-called
SCET$_{\rm I}$ in which virtuality of order $Q^2$ is integrated
out. Then we match, at the intermediate scale $Q^2(1-z)^2$, the
SCET DY cross section onto a product of SCET parton distribution
functions (PDF) \cite{collins}, giving rise to SCET$_{\rm II}$ in
which virtuality or order $Q^2(1-z)^2$ is integrated out. Here we
assume that the usoft interactions cancel as in the standard
Drell-Yan factorization which is valid in the kinematic region of
our interest \cite{fac}. The second matching yields the
infrared-safe hard scattering coefficient function. Then, the PDF
will run down in scale and be matched onto local twist-two
operators.

This paper is organized as follows. In section II we give the
basic notation, definitions and power counting of SCET. In section
III we perform the two-stage matching and write down the anomalous
dimensions. We compare our results for DY (as well as those of
DIS) with the full QCD calculations. In section IV we derive the
resummation (and exponentiation) of the large logs and compare our
results with those given in \cite{sterman8797,catani8991}. We
conclude the paper in Section V.

\section{SCET Preliminaries}

Soft-collinear effective field theory was invented to handle
processes with multiple momentum scales
\cite{bauer1,bauer2,feldman}. This is the case in DY and DIS when
final-state hadrons have an invariant mass much greater than
$\Lambda_{\rm QCD}$ but much smaller than the highest momentum
scale of the problem. In this section, we briefly describe some of
the concepts and notations in SCET. We choose
\begin{eqnarray}
n^{\mu}=\frac{1}{\sqrt2}(1,0,0,-1),\,\,\,\,
\bar{n}^{\mu}=\frac{1}{\sqrt2}(1,0,0,1),\,\,\,\, n^2=\bar{n}^2=0
 \ .
\end{eqnarray}
For a generic four-vector $l$ let $l^{\pm}\equiv
\frac{1}
{\sqrt2}(l^0 \pm l^3)$ and write
$l\equiv(l^+,l^-,l_\perp)$. Then
\begin{eqnarray}
l^+=n\cdot l,\,\,l^-=\bar{n}\cdot l, \,\,l^2=2l^+l^- +
l_{\perp}^2=2{n}\cdot l \bar{n} \cdot l + l_{\perp}^2 \ .
\end{eqnarray}
With this notation one has to trivially modify the SCET Feynman
rules given in \cite{bauer1}.

SCET describes the interaction of ``collinear'' quarks with
``collinear'' and/or ``soft'' gluons. If we have a fast-moving
quark or gluon in the $n$-direction then we assign to its momentum
components the following scaling: $p=(p^+,p^-,p_{\perp})\sim
Q(\lambda^2,1,\lambda)$ where $Q$ is some relevant hard scale and
$\lambda$ is a small parameter identified by considering the
different momentum scales available. A quark with such an
assignment is called $n$-collinear and is described by $\xi_n$
field and $n$-collinear gluon has $A^{\mu}_n$ field. Similarly,
for $\bar{n}$-collinear quark field we assign
$p=(p^+,p^-,p_{\perp})\sim Q(1,\lambda^2,\lambda)$ and is
described by $\xi_{\bar{n}}$ field. For the DY case, and in the
c.o.m. frame, we let the incoming quark be moving in the $+z$
direction (i.e $p_1^+$ is ``large'') and the incoming anti-quark
moving in the $-z$ (i.e. $p_2^-$ is ``large''); then these fields
are described by $\xi_{\bar{n}}$ and $\xi_n$ respectively. At the
scale $Q$, after integrating out virtuality of order $Q^2$, the
electromagnetic current in SCET$_{\rm I}$ is given by
\begin{eqnarray}
j^{\mu}&=& C(Q)\ \bar{\xi}_n W_n \gamma^{\mu}W_{\bar n}^\dagger
\xi_{\bar n} \ .
\end{eqnarray}
The coefficient $C(Q)$ has been calculated in \cite{manohar} and
will be presented in the next section. The $W_n$ and
$W^\dagger_{\bar n}$ are the familiar path-ordered Wilson lines
and are required to insure collinear gauge invariance of the
current operator \cite{bauer2,bauer3}. This
 Wilson line is
identical to the ``eikonal'' line introduced in \cite{collins}
\begin{eqnarray}
W_n(x)&=& {\rm P}\ {\rm exp}\left[ig\int_{-\infty}^{x} {\rm d}s\
{\bar n} \cdot A_n(s{\bar n})\right] \label{wilson} \ ,
\end{eqnarray}
where the covariant derivative is:
$D_{\mu}=\partial_{\mu}-igT^aA^a_{\mu}$. A ``soft'' gluon has the
momentum scale: $k=Q(\lambda^2,\lambda^2,\lambda^2)$. A SCET PDF
for a ${\bn}$-collinear field is defined as the matrix element of
the operator
\begin{eqnarray}
O_q(r^+) &=& \frac{1}{4\pi}  \int_{-\infty}^{\infty}\!\!\! {\rm
d}z e^{-i z r^+} \left[ \bar \xi_{\bar n}W_{\bar n}\right]\!\! (n
z) \not\! n \left[ W^\dagger_{\bar n}   \xi_{\bar n}\right] \!\!
\left( 0 \right) \ ,
\end{eqnarray}
between $\bn$-collinear quark fields. It is clear that for DY
process one needs to consider the $n$-collinear version of
Eq.~(5), i.e., interchanging ${\bn}\leftrightarrow n$ and taking
the matrix element between $n$-collinear fields.

In this work we use the Feynman gauge and the $\overline{\rm MS}$
scheme in $d=4-2\epsilon$. We will also use on-shell dimensional
regularization (DR) to regulate both the ultraviolet (UV) and the
infrared (IR) divergences. The choice of IR regulator in SCET is a
delicate matter \cite{salem}, and one has to make sure that the IR
regulator accurately generates the IR behavior of the full theory.
In Ref. \cite{manohar}, quark off-shellness was used as IR
regulator, however all the results can also be obtained by using
pure on-shell DR. This was checked explicitly. We will also follow
the normalization conventions for the DY and DIS cross sections
given in \cite{alta}. Thus the Born cross sections are
$\delta(1-z) \,\,{\rm and}\,\, \delta(1-x)$ respectively.

\section{Matchings At $Q^2$ and $Q^2(1-z)^2$ for DY: $p\overline p
\rightarrow l^+ l^- X$}

In this section, we derive the matching conditions for DY at $Q^2$
and $Q^2(1-z)^2$ at one-loop order. From these, we re-derive the
known one-loop coefficient function in SCET. To make the
discussion self-contained, we first review the result for DIS
process \cite{manohar}.

\subsection{Review of DIS in SCET}

Since the DY and DIS processes are related to each other, let us
review, briefly, how the analysis for DIS in SCET was performed.
Choosing the Breit frame, let the incoming quark be
${\bn}$-collinear, (i.e. moving in the $+z$ with large $p_1^+$ and
small $p_1^-$) and the outgoing quark be $n$-collinear, and let
$x\equiv {Q^2}/{2p_1\cdot q}$. The invariant mass of the final
hadronic state for DIS in the limit $x\rightarrow1$ is:
 $p_X^2\cong Q^2(1-x)$. Simple kinematic considerations show that
 for either collinear or soft gluon radiated into the final state:
 $p_X^2\cong Q^2\lambda^2+ O(\lambda^4)$ thus we identify:
 $\lambda^2\sim 1-x$.

 The Mellin transform of a function $f$ is defined
 as: $M_N[f(x)]=\int_0^1 {\rm d}x\ x^{N-1} f(x)$. $M_N$ is also known as
 ``$\rm{N}^{\rm{th}}$-moment'' of $f$. The limit $x\rightarrow1$
 corresponds to $N\rightarrow \infty$, thus one naturally,
 identifies, $1-x \sim 1/N$.

At scale $Q^2$, the final hadronic states may be effectively
considered as a massless jet moving in the $n$ direction. One
matches the full QCD current (i.e. only incoming and outgoing
quarks) onto SCET$_{\rm I}$ current given in Eq.(3). The matrix
element of the full QCD current in on-shell DR is given by
\begin{eqnarray}
\langle p_2 \vert j^{\mu} \vert p_1 \rangle &=& \gamma^{\mu} \left
\{ 1+ \frac{\alpha_s}{4\pi} C_F \left [ -\frac{2}{\eir^2} -
\frac{1}{\eir}\left(2 \ln \frac{\mu^2}{Q^2}+3\right) -{\ln}^2
\frac{\mu^2}{Q^2}- 3\ln \frac{\mu^2}{Q^2} -8 + \frac{ \pi^2}{6}
\right ]  \right\}
\end{eqnarray}
The $O(\alpha_s)$ virtual corrections to SCET current are
scaleless and vanish in pure DR, thus the matching condition at
$Q^2$ equals the finite part of Eq.~(6)
 \begin{eqnarray}
C_{\rm DIS}(\mu)&=& 1+ \frac{\alpha_s}{4\pi} C_F \left [ -{\ln}^2
\frac{\mu^2}{Q^2}- 3\ln \frac{\mu^2}{Q^2} -8 + \frac{ \pi^2}{6}
\right ] \ .
\end{eqnarray}
The anomalous dimension of SCET$_{\rm I}$ current satisfies
\begin{eqnarray}
\mu \frac{{\rm d}C_{\rm DIS}(\mu)}{{\rm d}\mu}&=& \gamma_1(\mu)\
C_{\rm DIS}(\mu) \ ,
\end{eqnarray}
which gives
\begin{eqnarray}
\gamma_1(\mu)&=& - \frac{\alpha_s}{4\pi}C_F \left[ 4 \ln
\frac{\mu^2}{Q^2} + 6 \right ] \ .
\end{eqnarray}

As the intermediate scale $Q^2(1-x)$, the final-state invariant
mass $p_X^2$ must be considered as ``large'', and the jet-like
final-state $n$-collinear hadronic modes can be integrated out.
This is done by matching a product of SCET$_{\rm I}$ currents onto
an effective theory (SCET$_{\rm II}$) where the $n$-collinear
quark is ``replaced'' by a soft Wilson line, thus one is led to
the matrix element of SCET$_{\rm II}$ quark operator given in
Eq.~(5) between quark states. This is very much like the standard
operator product expansion. If one computes the matching condition
at $Q^2(1-x)$ by using pure on-shell DR, then one has to consider
only the contribution of Feynman diagrams where a real gluon is
emitted into the final state, because the one-loop virtual
diagrams are scaleless and thus vanish.

If one computes the real-gluon emission diagrams in SCET$_{\rm
I}$, one finds the $O(\alpha_s)$ matching condition (coefficient
function) in \cite{manohar}
 \begin{eqnarray}
 \mathcal{M}_{\rm DIS}(x)&=& \frac{\alpha_s}{2\pi}C_F\ \theta(0 \le x \le 1) \Biggl\{ 2 \left[
\frac{{\rm ln}(1-x)}{(1-x)}\right]_+  + \Bigl[2 {\rm ln}\frac
{Q^2}{\mu^2}
 - \frac 32  \Bigr]\frac{1}{(1-x)_+}  \nonumber \\
 && + \Bigl[ \ln^2 \frac{Q^2}{\mu^2}
- \frac {3}{2}  \ln \frac{Q^2}{\mu^2}+  \frac {7}{2} -
\frac{\pi^2}{ 2} \Bigr] \delta(1-x) \Biggr\} \ .
\end{eqnarray}
The above result equals exactly to the finite part of the full QCD
calculation of the real gluon emission diagrams \cite{alta}, up to
terms that vanish in the large moment limit. Taking the product,
$C^2_{\rm DIS}(Q) [\delta(1-x)+\mathcal{M}_{\rm DIS}(x)]$ yields
the full QCD coefficient function. Therefore, the usual one-step
matching in DIS is divided into two steps, virtual and real
corrections, in SCET. This is also true in Drell-Yan process to be
discussed below.

\subsection{Drell-Yan Process}

In the Drell-Yan process, the invariant mass of the final hadronic
state is
\begin{eqnarray}
p_X^2&=& Q^2\left (1+\frac 1z -\frac 1 {x_1}- \frac 1 {x_2}\right
) \cong Q^2 \left [ (1-x_1)(1-x_2) \right ] \cong Q^2 \lambda^4 \
,
\end{eqnarray}
where
\begin{eqnarray}
z &=& \frac {Q^2}{s}, \,\,\, x_1= \frac{Q^2}{2p_1 \cdot q}, \,\,\,
x_2= \frac{Q^2}{2p_2 \cdot q} \ .
\end{eqnarray}
Since $1-z \sim \lambda^2$, $p_X^2 \sim Q^2(1-z)^2$. To integrate
out the final hadronic states, we match at the intermediate scale
$Q^2(1-z)^2$.

At the scale $Q^2$, the final-hadron states effectively have
$p_X^2=0$. We match the full QCD annihilation current: $q\bar
{q}\rightarrow \gamma^{*}$ onto the SCET$_{\rm I}$ current. The
matching condition can be obtained simply by analytically
continuing the full-QCD matrix element in Eq.~(6) from space-like
$q^2=-Q^2$ to time-like $q^2=Q^2$,
\begin{eqnarray}
\langle 0 \vert j^{\mu} \vert p_1\overline{p}_2 \rangle &=&
\gamma^{\mu} \left \{ 1+ \frac{\alpha_s}{4\pi} C_F \left [
-\frac{2}{\eir^2} - \frac{1}{\eir}\left(2 \ln
\frac{\mu^2}{Q^2}+3\right)\right.\right. \nonumber \\
&&\left.\left. -{\ln}^2 \frac{\mu^2}{Q^2}- 3\ln \frac{\mu^2}{Q^2}
-8 + \frac{ 7\pi^2}{6} \right ] \right \} \ .
\end{eqnarray}
The matching condition at $Q^2$ is thus
\begin{eqnarray}
C_{\rm DY}(\mu)&=& 1+ \frac{\alpha_s}{4\pi} C_F \left [ -{\ln}^2
\frac{\mu^2}{Q^2}- 3\ln \frac{\mu^2}{Q^2} -8 + \frac{7 \pi^2}{6}
\right ] \ .
\end{eqnarray}
The anomalous dimension is the same as Eq.~(9). It should be noted
that $\eir$-dependent terms in Eq.~(13) equal the negative of the
$\euv$ terms in the effective theory. This is so since, as
mentioned earlier, the SCET virtual contributions vanish in pure
on-shell DR due to cancellation of the form
$\frac{1}{\euv}-\frac{1}{\eir}$, and because the IR behavior of
the effective theory must match that of the full QCD. Thus the UV
counter-term for the current in the effective theory equals
\begin{eqnarray}
{\rm c.t.}&=& \frac{\alpha_s}{2\pi} C_F \left [
-\frac{1}{\epsilon^2} - \frac{3}{2\epsilon} - \frac{1}{\epsilon}
\ln \frac{\mu^2}{Q^2} \right ] \ .
\end{eqnarray}

Below $Q^2$, the DY process is described by SCET$_{\rm I}$.
Because the virtual diagrams vanish in on-shell DR, we need only
consider real emission diagrams given in Figs. 1 and 2, where the
cut lines go through the real gluon in the final state. Let us
take, in the c.o.m. frame, the momentum $k^{\mu}$ of the emitted
gluon as:
\begin{eqnarray}
 k^{\mu}=(\vert k \vert,...\vert k \vert \cos \theta)\ ,
\end{eqnarray}
and introduce the Mandelstam variables (besides $s$)
\begin{eqnarray}
t &=& -\frac{Q^2}{z}(1-z)(1-y),\,\,\,\,\,\, u=-\frac{Q^2}{z}(1-z)y
\ ,
\end{eqnarray}
where: $y=\frac12 (1+ \cos \theta)$. The calculation of the
contributions from the above diagrams is performed by taking the
square of the amplitude for emitting a real gluon and then
integrating over the phase space for the production of a massive
photon \cite{alta}
\begin{eqnarray}
{\rm PS}&=& \frac{1}{8\pi}\left(\frac{4\pi}{Q^2} \right
)^{\epsilon} \frac{1}{\Gamma (1-\epsilon)}z^\epsilon
(1-z)^{1-2\epsilon} \int_{0}^{1}{\rm d}y (y(1-y))^{-\epsilon} \ .
\end{eqnarray}
Diagrams (c) in Fig.1 and (a)-(b) in Fig.~2 are zero due to
$n^2=\bn^2=0$. Diagrams (a) and (b) have the same contribution as
their full QCD counterparts. The sum of the three remaining
contributions completes the total result to that of full QCD which
can be found in, e.g, \cite{alta}. We perform the phase space
integration for the sake of completeness, keeping only singular
terms in the limit $z\rightarrow 1$,
\begin{figure}[t]
\begin{center}

\includegraphics{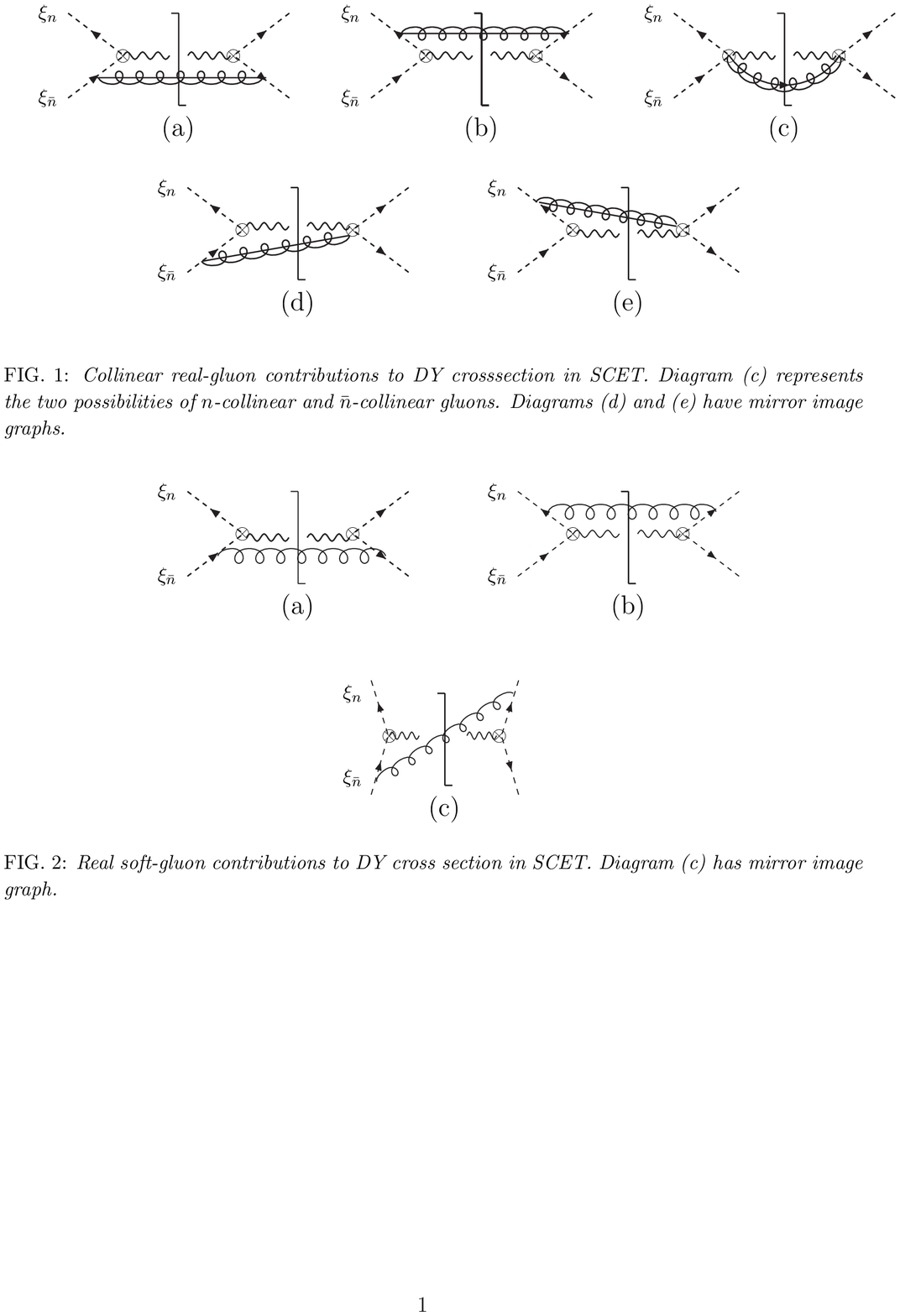}
\end{center}
%\vskip -0.7cm
 %\caption{\it Real soft-gluon contributions to DY
%cross section in SCET. Diagram (c) has mirror image graph. }
\end{figure}

\begin{eqnarray}
\mathcal{I}_{\rm DY}(z)&=& \frac{\alpha_s}{2\pi}C_F\ \theta(0 \le
z \le 1)\left (\frac{\mu^2 e^{\gamma}}{Q^2}\right)^\epsilon
\frac{z^{\epsilon}(1-z)^{1-2\epsilon}}{\Gamma(1-\epsilon)}\int_0^1
{\rm d}y [y(1-y)]^{-\epsilon}\times
\frac{2}{y(1-y)}\frac{1}{(1-z)^2} \nonumber \\
&=& \frac{\alpha_s}{2\pi}C_F\ \theta(0 \le z \le 1)\times 2 \left
[ \frac{1}{\epsilon^2}\
\delta(1-z)-\frac{2}{\epsilon}\left(\frac{1}{1-z}\right)_{+} +
4\left(\frac{\ln(1-z)}{1-z}\right)_{+} \right ]
\nonumber \\
&&\times \left [1+ \epsilon \ln \frac{\mu^2}{Q^2} + \epsilon^2
\left( \frac{1}{2} \ln^2 \frac{\mu^2}{Q^2} -
\frac{\pi^2}{4}\right)\right]\nonumber \\
&=&  \frac{\alpha_s}{2\pi}C_F\ \theta(0 \le z \le 1) \left [
 \frac{2}{\epsilon^2}\ \delta(1-z) + \frac{2}{\epsilon} \ln
\frac{\mu^2}{Q^2} \delta(1-z) - \frac{4}{\epsilon} \left (
\frac{1}{1-z}\right )_{+}\right. \nonumber \\
&& \left. +8\left ( \frac {\ln (1-z)}{1-z}\right )_+
   -4\ln \frac{\mu^2}{Q^2} \left ( \frac {1}{1-z} \right )_{+} +
\delta(1-z) \left ( \ln^2 \frac{\mu^2}{Q^2} -
\frac{\pi^2}{2}\right )  \right ] \ . \label{result}
\end{eqnarray}
In the above calculation we ignored the contribution from ${\ln
z}/(1-z)$ since this function is regular in the limit
$z\rightarrow 1$ and the corresponding Mellin moments vanish at
large $N$. As expected, the result is identical to the full QCD
calculation of the real gluon contributions to the DY differential
cross section in the limit $z\rightarrow1$ \cite{alta}. The
$\epsilon$-terms in Eq.~(19) are a combination of UV and IR
contributions.

To see how to interpret these terms, we need to consider the PDF
given in Eq.~(5). In pure on-shell DR, the $O(\alpha_s)$
corrections to the unrenormalized parton distribution vanish due
to scaleless integrals. The UV-behavior of the SCET PDF is given
by \cite{manohar}: $\frac{\alpha_s}{2\pi\euv}P_{q\leftarrow q}(z)$
where in the limit $z\rightarrow1$
\begin{eqnarray}
 P_{q\leftarrow q}&=& C_F \left [ \frac{2}{(1-z)_+} +
\frac{3}{2} \delta(1-z) \right ]\theta (0\leq z \leq 1) \ ,
\end{eqnarray}
which is the well-known Altarelli-Parisi splitting kernel. Thus
the renormalized parton distribution is
$-\frac{\alpha_s}{2\pi\eir}P_{q\leftarrow q}(z)$. Writing the
$\epsilon$-terms in Eq.~(19) as
\begin{eqnarray*}
\frac{\alpha_s}{2\pi}C_F\ \theta(0 \le z \le 1) \left [
 \left(\frac{2}{\epsilon^2} + \frac{3}{\epsilon} + \frac{2}{\epsilon} \ln
\frac{\mu^2}{Q^2}\right) \delta(1-z) - \frac{4}{\epsilon} \left (
\frac{1}{1-z}\right )_{+} - \frac{3}{\epsilon}\delta(1-z) \right ]
\ .
\end{eqnarray*}
Taking into account the counter-term for the SCET current,
Eq.~(15), and matching the DY cross section onto a product of two
PDFs, all the $\epsilon$-dependent terms cancel.

The finite part of Eq. (\ref{result}) yields $O(\alpha_s)$
matching condition (coefficient function)
\begin{eqnarray}
\mathcal{M}_{\rm DY}(z)&=& \frac{\alpha_s}{2\pi}C_F\ \theta(0 \le
z \le 1)\left [ 8\left ( \frac{\ln(1-z)}{1-z} \right )_+ -4\ln
\frac{\mu^2}{Q^2} \left ( \frac{1}{1-z} \right )_+ \right.
\nonumber \\
&& \left. + \delta(1-z) \left ( \ln^2 \frac{\mu^2}{Q^2} -
\frac{\pi^2}{2} \right ) \right ] \label {M_dy} \ .
\end{eqnarray}

The moments of the matching condition at $Q^2(1-z)^2$, Eq.~(21),
are
\begin{eqnarray}
\mathcal{M}_{\rm DY}(N)&=& \frac{\alpha_s}{2\pi}C_F \left [ \ln^2
\frac{{\overline N}^2 \mu^2}{Q^2}+ \frac{\pi^2}{6} \right ] \ ,
\end{eqnarray}
where $\overline N = N \exp(\gamma_E)$ and $\gamma_E$ is the Euler
constant. In order to minimize the large logs we choose
$\mu=Q/{\overline N}$. This is different from the DIS case where
at the intermediate scale we set $\mu=Q/{\sqrt {\overline N}}$.
Thus for DY we identify $1-z\sim 1/N$. At this stage it is
appropriate to clarify the following point: the above calculation
demonstrates the all order factorization theorem
\cite{fac,Collins} of the DY process to first order in $\alpha_s$,
i.e, all the IR divergences are contained into a product of two
PDFs defined in Eq.(5).

Below the scale $Q^2(1-z)^2$, we have SCET$_{\rm II}$ in which the
all hard scales have been integrated out. The running of the
moments of the PDF with $\mu$ is governed by the well-known
anomalous dimension $-\gamma_{2,N}$ \cite{manohar,field} where
\begin{eqnarray}
\gamma_{2,N}&=& -\frac{\alpha_s}{2\pi} C_F \left[
4\sum_{j=2}^{N}\frac{1}{j} - \frac{2}{N(N+1)} +1 \right] \ ,
\end{eqnarray}
In \cite{manohar} it was shown, to first order in $\alpha_s$, that
the running of the PDF comes from Fenyman diagrams where only
collinear quarks and gluons interact and the soft contribution (or
usoft- in the terminology of Ref.\cite{manohar}) vanishes. This is
of-course consistent with the factorization theorem of DY which we
assume that it holds to all orders in perturbation theory.\\
 In the large $N$ limit we have
\begin{eqnarray}
\gamma_{2,N}&\cong& -\frac{\alpha_s}{2\pi} C_F \left[4 \ln
{\overline N}-3 \right ] \ .
\end{eqnarray}
At the lowest scale $\mu_0$, one identifies the PDF moments as the
matrix elements of local twist operators $A_N(\mu_0)$
\cite{bardeen}.

\section{Resummation and Exponentiation}

By now, we have obtained all the ingredients we need to write down
an expression for the moments of the DY differential cross section
$(\frac{{\rm d}\sigma_{\rm DY}}{{\rm d}Q^2})$. By taking the
square of the current matching coefficient at $\mu^2=Q^2$, running
down the scale with anomalous dimension $2\gamma_1$ to
$\mu^2=Q^2/{\overline N}^2$, we then multiply with the moments of
matching condition into the product of the matrix element of PDF.
Then we run down the scale to $\mu\simeq\Lambda_{\rm QCD}$ with
$2\gamma_2$ and match onto the moments of local twist-two
operators. In summary, one gets
 \begin{eqnarray}
 \left (\frac{{\rm d}\sigma_{\rm DY}}{{\rm d} Q^2}\right )_N &=& C^2_{\rm DY}
 (Q)e^{-I_{\rm DY1}}\left( 1+
 {\mathcal M_{\rm DY}(N,\mu=Q/{\overline N})}\right ) e^{-I_{\rm
 DY2}}A_N^2(\mu_0)\ ,
 \label{final}
 \end{eqnarray}
where
\begin{eqnarray}
I_{\rm DY1} = \int_{Q/{\overline N}}^{Q} \frac{{\rm
d}\mu}{\mu}2\gamma_1(\mu); ~~~~~  I_{\rm DY2} =
\int_{\mu_0}^{Q/{\overline N}} \frac{{\rm
d}\mu}{\mu}2\gamma_2(\mu) \ ,
\end{eqnarray}
where $\gamma_1(\alpha_s, Q^2/\mu^2)$ is the anomalous dimension
of the SCET current and $\gamma_2(\alpha_s, N)$ is the anomalous
dimension of the twist-two operator.

The above can be compared with the DIS result,
\begin{equation}
   F_{2N}(Q^2) = C^2_{\rm DIS}(Q) e^{-I_{\rm DIS1}}\left(1+
 {\mathcal M_{\rm DIS}(N,\mu=Q/\sqrt{{\overline N}})}\right ) e^{-I_{\rm DIS2}}A_N(\mu_0)
\end{equation}
where
\begin{eqnarray}
I_{\rm DIS1} = \int_{Q/\sqrt{{\overline N}}}^{Q} \frac{{\rm
d}\mu}{\mu}2\gamma_1(\mu); ~~~~~  I_{\rm DIS2} =
\int_{\mu_0}^{Q/{\sqrt{\overline N}}} \frac{{\rm d}\mu}{\mu}
\gamma_2(\mu) \ .
\end{eqnarray}
Let us define a physical observable which is a ratio of the
moments between DY and DIS squared,
\begin{eqnarray}
    \Delta_N &=& \frac{(d\sigma_{DY}/dQ^2)_N}{F_{2N}^2} \ .
\end{eqnarray}
At one-loop order, one has
\begin{equation}
    \Delta_N = 1 + \frac{\alpha_s}{2\pi} C_F
            \left( 2\ln^2\overline{N} - 3\ln\overline{N} + 1  +
            \frac{5}{3}\pi^2\right) \ ,
\end{equation}
which is consistent with the known result. When summing over
higher-order corrections of form $\alpha_s^{n}\ln^m\overline N$
with $m\le 2n$, $\Delta_N$ can be expressed in terms of an
exponential form, $\exp(f)$, with $f = \ln\overline N
f_{-1}(\alpha_s \ln \overline N) + f_0(\alpha_s \ln \overline N)
+\alpha_s f_1(\alpha_s\ln \overline N) + ...$. The first term
yields the leading double-logarithm resummation, and the second
gives the next-to-leading logarithm resummation (NLL), etc. In the
present calculation, we can have an accuracy up to NLL.

To NLL resummation, $\Delta_N$ can be expressed as
\begin{equation}
      \ln \Delta_N =  \int^Q_{Q/\sqrt{\overline N}} 2\gamma_1(\alpha_s(\mu))
      \frac{d\mu}{\mu} + \int^{Q/\sqrt{\overline N}}_{Q/\overline
      N}
      2\left[\gamma_1(\alpha_s(\mu))-\gamma_2(\alpha_s(\mu))\right]\frac{d\mu}{\mu}\
      ,
\end{equation}
where the anomalous dimensions have the form
\begin{eqnarray}
   \gamma_1 &=& \Gamma_1 \ln Q^2/\mu^2  + \Gamma_2 + \Gamma_3/2\ ,  \nonumber \\
   \gamma_2 &=& 2\Gamma_1 \ln \overline N + \Gamma_3 \ ,
\end{eqnarray}
to all orders in perturbation theory. To NLO, one has $\Gamma_3 =
2\Gamma_2$, and the above expression can be expressed as
\begin{equation}
  \ln  \Delta_N = 2 \int^1_{{\overline N}^{-1}} \frac{dy}{y}
    \left[ \int^{\sqrt{y}Q}_{yQ}
    2\Gamma_1(\alpha_s(\mu))\frac{d\mu}{\mu}
      +\Gamma_2(\alpha_s(\sqrt{y}Q))\right] \ .
\end{equation}
This result does not have an infrared renormalon or Landau
singularity because the expression is manifestly valid only when
the scale $Q/\overline N$ is much larger than $\Lambda_{\rm QCD}$.
If one takes the limit $N\rightarrow \infty$ with a fixed $Q$, the
above result can also be expressed as
\begin{equation}
   \Delta_N = \exp\left[-2\int^1_0 dz
   \frac{z^{N-1}-1}{1-z}
  \left( \int^{Q^2}_{Q^2(1-z)}
     \Gamma_1(\alpha_s(k^2))\frac{dk^2}{k^2}
      + \Gamma_2\left(\alpha_s((1-z)Q^2)\right)\right)\right]
\end{equation}
which is the same as what has been obtained by Catani and
Trentadue \cite{catani8991}, and Sterman \cite{sterman8797}. This
expression has the infrared renormalon problem because the
integral covers the soft momentum region where the strong coupling
constant is ill-defined.

It is instructive to recall how the resummation in the large-$x$
limit is achieved in Sterman's approach \cite{sterman8797}. The DY
cross section and DIS structure function can be factorized in
$x\rightarrow 1$ limit into a product of \emph{hard parts}, a
\emph{soft part}, and \emph{collinear parts}. Different parts may
be gauge-dependent, but the product is not. The renormalization
group equations for individual parts can be derived
straightforwardly.

The hard part contains lines with momentum of order $Q$ and is
included in the first stage matching in SCET approach. The soft
part is defined as the matrix element of Wilson lines and is
infared finite (for a definition in SCET, see \cite{chay}). This
soft contribution comes in fact from gluon lines of virtuality of
order $Q^2(1-x)$, which can be calculated in perturbation theory
when $Q^2(1-x)\gg \Lambda^2_{\rm QCD}$ \cite{sterman8797}. In SCET
approach, this contribution is taken into account by the second
stage matching. The collinear parts are either special parton
distributions or jet functions or both. They depend on a rapidity
cut-off and obey an evolution equation in rapidity. This equation
is quite general, and when solved, contains the resummed double
logarithms. When $Q^2(1-x)\gg \Lambda^2_{\rm QCD}$, the special
parton distriubtions can be factorized in terms of ordinary parton
distributions, whereas the jet functions can be calculated
entirely in pQCD.

\section{conclusion}

Following a previous work on summing large logarithms as
$x\rightarrow 1$ in DIS, we use the same SCET formalism to study
the resummation in DY. The steps are quite similar and the result
is shown in Eq. (\ref{final}). The result is consistent with the
known result, but is free of the Landau singularity.

We thank A. Manohar and G. Sterman for useful discussions, and the
work is partially supported by the U. S. Department of Energy via
grant DE-FG02-93ER-40762.

\end{document}